\documentclass[12pt]{article}
\usepackage{algorithm}
\usepackage{algorithmic}
\usepackage{tabularx}
\usepackage{booktabs}
\usepackage{amssymb}
\usepackage{multirow}
\usepackage{graphicx}
\usepackage[english]{babel}
\usepackage[letterpaper,top=3cm,bottom=3cm,left=3cm,right=3cm,marginparwidth=1.75cm]{geometry}
\usepackage{cite}
\usepackage{amsmath}
\usepackage{graphicx}
\usepackage[colorlinks=true, allcolors=blue]{hyperref}
\usepackage{authblk}
\usepackage[numbers]{natbib}
\usepackage{graphicx}
\usepackage{cleveref}
\title{XTransCT: Ultra-Fast Volumetric CT Reconstruction using Two Orthogonal X-Ray Projections for Image-guided Radiation Therapy via a Transformer Network}
\author[1]{Chulong Zhang}
\author[1]{Lin Liu}
\author[1]{Jingjing  Dai}
\author[1]{Xuan Liu}
\author[1]{Wenfeng He}
\author[1]{Yinping Chan}
\author[1]{Yaoqin Xie}
\author[2,*]{Feng Chi}
\author[1,*]{Xiaokun Liang}
\affil[1]{Shenzhen Institute of Advanced Technology, Chinese Academy of Sciences, Shenzhen, 518055 Guangdong, China}
\affil[2]{State Key Laboratory of Oncology in South China, Guangdong Provincial Clinical Research Center for Cancer, Sun Yat-sen University Cancer Center, 510060, China.}
\affil[*]{Email:  \href{mailto:xk.liang@siat.ac.cn}{xk.liang@siat.ac.cn} / \href{mailto:chifeng@sysucc.org.cn}{chifeng@sysucc.org.cn}}                     
\begin{document}
\maketitle

\begin{abstract}

\textbf{Objective:} The aim of this study was to reconstruct Volumetric Computed Tomography (CT) images in real-time from ultra-sparse two-dimensional X-ray projections, facilitating easier navigation and positioning during image-guided radiation therapy.

\textbf{Approach:} Our approach leverages a voxel-sapce-searching Transformer model to overcome the limitations of conventional CT reconstruction techniques, which require extensive X-ray projections and lead to high radiation doses and equipment constraints.

\textbf{Main Results:} The proposed XTransCT algorithm demonstrated superior performance in terms of image quality, structural accuracy, and generalizability across different datasets, including a hospital set of 50 patients, the large-scale public LIDC-IDRI dataset, and the LNDb dataset for cross-validation. Notably, the algorithm achieved an approximately 300$\%$ improvement in reconstruction speed, with a rate of 44 ms per 3D image reconstruction compared to former 3D convolution-based methods.

\textbf{Significance:} The XTransCT architecture has the potential to impact clinical practice by providing high-quality CT images faster and with substantially reduced radiation exposure for patients. The model’s generalizability suggests it has the potential applicable in various healthcare settings. 

\end{abstract}




\maketitle

\section{Introduction}

Image-guided radiation therapy (IGRT) is an increasingly essential component of cancer radiation therapy, where various imaging techniques are used to the planning and execution of treatment process \cite{xing2006overview}. By utilizing imaging technologies, such as computed tomography (CT) scans, IGRT enables the precise targeting of tumors while minimizing the radiation dose delivered to health surrounding tissue. IGRT enables the measurement of changes in tumor size, position, and shape, leading to more accurate radiation delivery and the ability to escalate the radiation dose to the tumor. This approach also reduces variability in delivered doses across patient population, thereby enhancing the interpretation of clinical trial outcomes.

\begin{figure}
\centering
\includegraphics[width=0.6\textwidth]{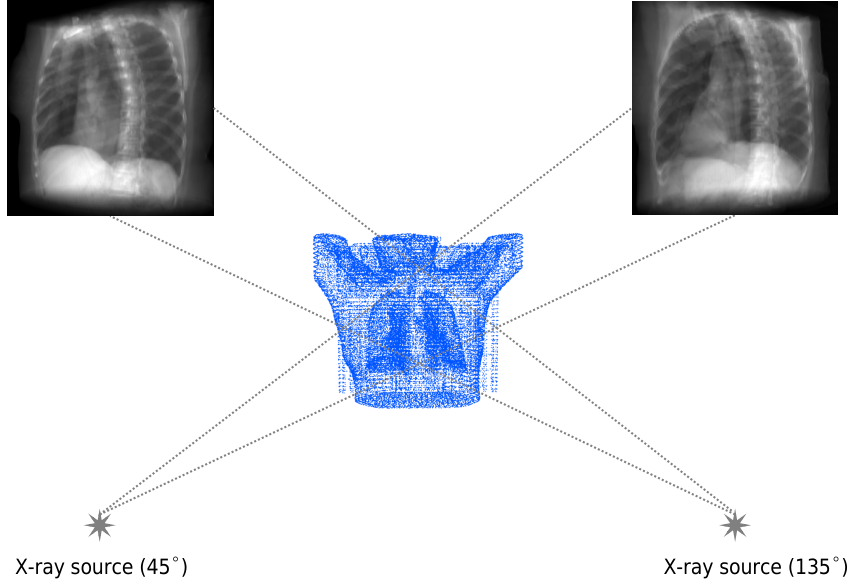}
\caption{Dual X-ray setup. Two X-rays irradiate the patient, and their real-time fusion with CT scans enables precise patient positioning while accounting for potential misalignments. In this study, we adopt a similar strategy, setting our two ray source angles at 45° and 135°.
}
\label{figure1}
\end{figure}

In IGRT, our primary goal has been to achieve lower radiation dose \cite{zhang2018sparse}, minimal hardware burden, and higher real-time performance. Accordingly, we aimed to achieve intraoperative positioning and navigation with the fewest possible X-ray images. One such system that has been proposed is Brainlab's ExacTrac X-Ray six-dimension system \cite{jin2008use}. This system, are shown in Figure \ref{figure1}, comprises only two oblique x-ray imager sets to determine patient position and acquire high-quality radiographs for patient position verification and adjustment. This system relies on two oblique x-ray images to register the resulting two-dimension (2D) images with the three-dimension (3D) CT images for  Radiation Therapy guidance. However, it is limited in its ability to achieve position verification and adjustment for patients with limited degrees of freedom (i.e., three or six degrees of freedom).

The advent of deep learning has made it possible to perform tasks previously limited to 3D images using ultra-sparse X-ray images. One approach involves 2D / 3D registration \cite{dong20232d,zhang20212d}, while another directly reconstructs 3D images from ultra-sparse 2D images. For instance, X2CT-GAN \cite{ying2019x2ct} is an end-to-end CT image generation method that utilizes a 2D convolutional neural network (CNN), a 3D CNN, and generative adversarial networks. These algorithms process feature blocks formed after passing 2D images through 2D CNNs and perform the reconstruction or alignment tasks end-to-end in 3D CNNs. However, these methods experience a significant efficiency loss as they only use two sparse information-containing 2D images. Recently, SimpleRecon \cite{sayed2022simplerecon} proposed an alternative 3D image reconstruction approach that does not require 3D convolution, significantly improving image reconstruction speed.

Generating three-dimensional images from two-dimensional X-ray images offers a straightforward solution for acquiring a patient's positional information in space during radiation therapy. However, methods relying exclusively on 2D CNNs to aggregate information from images have limited accuracy in processing such data, often resulting in subpar performance. Combining 2D and 3D CNNs to process 2D images could address this issue, but it frequently leads to information redundancy. Thus, designing a method that effectively and efficiently integrates both 2D and 3D information without redundancy is imperative.

Our work draws inspiration from the Neural Radiance Field (NeRF) \cite{mildenhall2021nerf}, which represents natural images in 3D space using $(x,y,z,\theta,\phi)$ coordinates, where $(x,y,z)$ denote the spatial coordinates and $(\theta,\phi)$ indicate the viewing direction. The reconstruction process maps $(x,y,z,\theta,\phi)$ to $(r,g,b,\sigma)$, with $(r,g,b)$ representing color and $\sigma$ signifying density. In contrast to other imaging modalities, medical imaging scenes are simpler. We assume that $(x,y,z)$ represents the spatial coordinates, while $\theta$ denotes grayscale. Consequently, mapping from $(x,y,z)$ to $\theta$ is sufficient for representing the reconstruction process. To accomplish this, we introduce a Voxel-Space Searching Transformer, which is built upon the structure of the transformer. Here, we consider $(x,y,z)$ as queries and decode the corresponding $\theta$ in the transformer network.

We present a novel approach for end-to-end reconstruction of X-rays into CT scans, allowing for any number of X-rays. However, we suggest using biplanar X-rays as a practical solution. A single X-ray image lacks the necessary information to generate a reliable CT image, while employing multiple X-ray images would necessitate additional equipment and increase the burden on the imaging system. For instance, Brainlab's ExacTrac product \cite{jin2008use} utilizes two X-ray images for correcting and positioning patients during surgical procedures.

Due to challenges in obtaining X-ray projection data in real-world settings, CT data has emerged as a more convenient alternative. Consequently, digitally reconstructed radiograph (DRR) technology is employed to simulate X-rays for 2D / 3D tasks, as demonstrated in prior studies \cite{dong20232d,ying2019x2ct,zhang20212d,wang2017pulmonary,ge2022x}. Our task involves reconstructing a CT image from two X-ray images, specifically at 45° and 135° angles. As our goal is focused on surgical guidance rather than clinical diagnosis, we assess both the image quality and structural integrity of our generated images relative to the segmentation labels. Moreover, considering the need for real-time surgical navigation during procedures, we also evaluate inference time as a critical performance metric.

Our significant technical contributions include:

$\bullet$ Proposing a novel voxel-space searching transformer for real-time volumetric CT image reconstruction, representing the first attempt in this domain.

$\bullet$ Our proposed method eliminates the need for 3D convolution, resulting in a considerably faster algorithm. In particular, our approach is several times quicker than methods employing 3D CNNs.

$\bullet$ We assessed our method in terms of speed, image quality, and structural accuracy, achieving state-of-the-art performance.

\begin{figure*}
\centering
\includegraphics[width=1\textwidth]{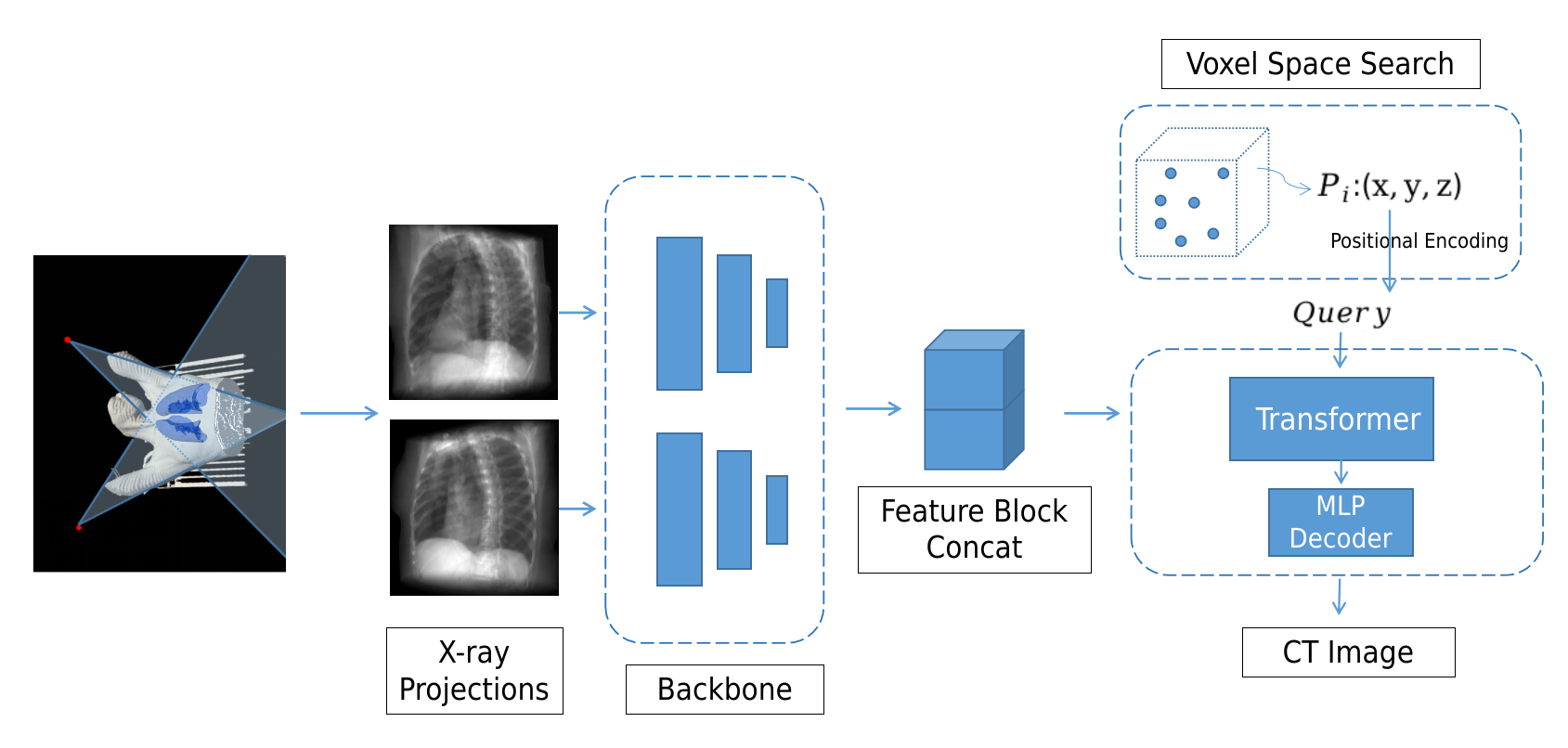}
\caption{
Framework of this study. We employ the DRR method to generate X-ray images simulated from CT scans. Our objective is to reconstruct the original CT scan using two X-ray images. Initially, we input these two DRR images into the backbone network to obtain feature blocks of size $8\times16\times512$. Next, we feed these feature blocks into the Transformer encoder. We then use a voxel space search module that traverses $\mathcal{P}:(x,y,z),\mathcal{P} \in {[0,1]}^{3}$ as queries and inputs them into the Transformer decoder. The Transformer decoder produces a 256-dimensional feature, representing the grayscale value $\theta$ corresponding to $\mathcal{P}$. By arranging the value of $\theta$ according to the order of $\mathcal{P}$, we can form the predicted complete 3D image $I^\prime$.
}
\label{figure2}
\end{figure*}

\section{Related Works}
\noindent\textbf{Image-guided Radiation Therapy} 
IGRT is a contemporary radiation therapy technique that delivers high-dose radiation with exceptional precision to the target tumor site \cite{xing2006overview}. The IGRT procedure depends on various imaging technologies, including X-rays, CT, magnetic resonance imaging, and positron emission tomography for accurate visualization of the tumor and surrounding tissue.

In the era of deep learning, numerous deep learning methods have been applied to IGRT, particularly in image registration and tumor tracking. However, this study focuses on tracking and localizing a patient's internal state during the radiation therapy process. Commercially available products, such as ExacTrac, have been used to achieve a limited range of motion by employing two X-rays. Currently, researchers commonly utilize 2D / 3D registration or image generation techniques to accomplish this task.

\noindent\textbf{2D / 3D tasks in medical image} 2D / 3D tasks in medical imaging can be broadly categorized into model-driven and deep learning-based approaches \cite{dong20232d}. Model-driven methods involve motion modeling techniques, such as Principal Component Analysis, Feature Descriptor model-agnostic deformation methods, and elastic registration methods based on biomechanical models for 2D / 3D registration \cite{ref13,ref19}. However, these techniques generally require iterative optimization of registration images, resulting in slow registration speeds that do not fulfill the real-time requirements for intraoperative IGRT.

Data-driven models employing end-to-end learning strategies have become increasingly important for 2D / 3D applications, owing to their ability to reduce reference time. Data-driven techniques for 2D / 3D can be classified into three categories: 2D / 3D registration, 2D / 3D segmentation, and 2D / 3D reconstruction. Data-driven registration methods comprise various techniques, such as CNN regression methods for 2D / 3D rigid registration \cite{ref18}, projection space transformation for spatial transformation \cite{ref15}, point-of-interest networks for 2D / 3D rigid registration under multiple views \cite{ref16}, and deep learning-based methods for non-rigid registration \cite{dong20232d,ref25}.

\noindent\textbf{Methods of spatial coordinate mapping} DeepSDF \cite{park2019deepsdf} and NeRF \cite{mildenhall2021nerf} are two prominent algorithms used for point cloud reconstruction. The fundamental principle of DeepSDF is to employ deep neural networks to predict continuous SDF values from given points. In other words, it establishes a mapping between point coordinates and SDF, enabling precise representation of an object's geometry. COTR \cite{jiang2021cotr}, similarly inspired by DeepSDF, utilizes coordinate mappings for image matching tasks. Conversely, NeRF focuses on mapping coordinates to RGB values and densities. In this paper, we draw inspiration from these state-of-the-art algorithms to propose a transformer-based approach addressing the challenge of mapping spatial coordinates to corresponding voxels in medical images. To the best of our knowledge, our proposed method is the first to address this issue.

\noindent\textbf{Attention mechanisms} The attention mechanism is an essential feature of neural networks, allowing them to selectively focus on specific aspects of input data. The concept of "hard" attention mechanisms, which utilize a differentiable sampler, was initially introduced by Spatial Transformers \cite{jaderberg2015spatial} and has since experienced further development. In contrast, Transformers \cite{vaswani2017attention} introduced the notion of "soft" attention mechanisms, which are widely used in natural language processing and are beginning to gain traction in visual tasks. The Transformer architecture has demonstrated impressive performance in computer vision, with examples such as DETR \cite{carion2020end} for object detection and ViT \cite{dosovitskiy2020image} for image recognition. Recently, the Swin Transformer \cite{liu2021swin} has achieved state-of-the-art results in various visual scene understanding tasks, gaining widespread adoption in areas such as image classification, object detection, and segmentation.

\section{Method}

In this paper, we first formulate the problem statement in Section 3.1. Next, we provide a detailed description of our network framework (\ref{figure2}) in Section 3.2, followed by the introduction of our accelerated voxel space search strategy in Section 3.3.

\subsection{Problem formulation}

\begin{figure}[h] \centering 
\includegraphics[height=6cm]{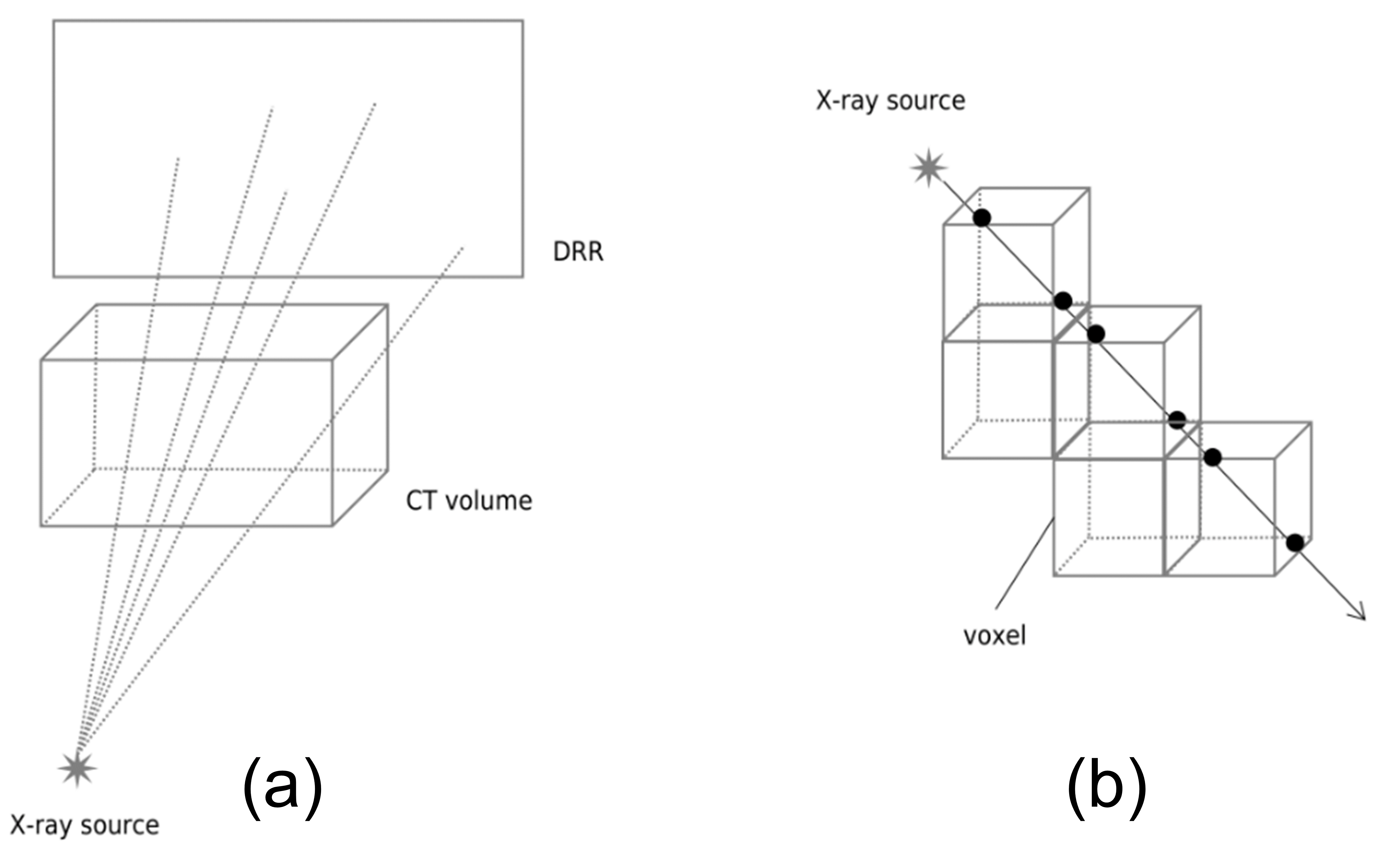} 
\caption{(a) X-rays passing through an object, resulting in a single projection on a plane. (b) targeted X-ray beam passing through a voxel.} 
\label{figure3} 
\hfill 
\end{figure}

DRR is a technique that simulates ray images by projecting 3D volume images onto a 2D image plane. Current DRR reconstruction algorithms primarily employ ray casting, a well-established method in computer graphics and 3D visualization. This technique mimics the process of X-rays passing through the human body, attenuating as they are absorbed by tissue, and generating DRR images using the principle of ray projection.

Initially, a virtual point source of X-rays was simulated to represent a conventional X-ray source, as illustrated in Figure \ref{figure3}(a). Multiple virtual X-rays were then emitted from this point source, passing through the three-dimensional dataset and projecting onto a panel perpendicular to the central axis of the X-ray beam. The projection points of all X-rays on the panel corresponded to the pixels of the DRR image, as depicted in Figure \ref{figure3}(b). During the projection process, X-rays traversed the three-dimensional dataset at predetermined intervals, and at each CT slice, the X-ray attenuation was simulated using an algorithm to obtain a CT value. If the intersection of the X-ray and the slice did not coincide with a pixel, interpolation algorithms were employed to estimate the CT value at that location. Finally, the CT values of all intersections on each X-ray were summed, and the accumulated value was converted into a grayscale image through appropriate transformations to obtain the DRR image.

The goal of our task is to establish a mapping between the two point light sources generated for performing DRR on the 3D image volume $I$, with a size of 128$\times$128$\times$128. Imaging is performed at angles of 45° and 135°, resulting in the production of two DRR images, $I_1$ (256$\times$256) and $I_2$ (256$\times$256). Assuming a 3D point $\mathcal{P}:(x,y,z),\mathcal{P} \in {[0,1]}^{3}$ within the voxel space corresponds to a gray value $\theta$, a straightforward approach is to use $I_1$ and $I_2$ to obtain all the gray values for each $\mathcal{P}$. This process results in a mapping between $(I_x,I_y)$ and $I$, allowing for the reconstruction of $I$ using both DRR images. Therefore, our objective function is:

\begin{equation}
    \begin{aligned}
        \underset{\boldsymbol{\Phi}}{\arg \min } \underset{\left( \boldsymbol{I}, \boldsymbol{I_1}, \boldsymbol{I_2}\right) \sim \mathcal{D} }{\mathbb{E}} \left\|\boldsymbol{\theta}-\mathcal{F}_{\boldsymbol{\Phi}}\left(\boldsymbol{\mathcal{P}} \mid \boldsymbol{I_1}, \boldsymbol{I_2}\right)\right\|_{2}^{2}
    \end{aligned}
\end{equation}

where $\mathcal{D}$ denotes the range of $\left( \boldsymbol{\mathcal{I}},\boldsymbol{I_1}, \boldsymbol{I_2}\right) $ under the dataset features and $\Phi$ is the parameter of the network.

\subsection{Network Architecture}

In this study, we present a novel framework for reconstruction using a 2D CNN and Transformer architecture. Our approach differs from traditional methods that employ 3D CNNs. Specifically, we utilize the spatial coordinates in voxel space as queries and decode the unknown voxel value for each coordinate.

Initially, we input two DRR images into the backbone to generate a feature block of size $8\times16\times256$, which is subsequently passed to the Transformer encoder. Next, we integrate a voxel space search module, where we explore all points $\mathcal{P}:(x,y,z),\mathcal{P} \in {[0,1]}^{3}$ as queries input to the Transformer decoder. The resulting 256-dimensional feature output from the Transformer decoder represents the gray value $\theta$ corresponding to $\mathcal{P}$. By arranging $\theta$ in the sequence corresponding to $\mathcal{P}$, we obtain the predicted complete 3D image $I^\prime$. The loss function is defined as:

\begin{equation}
\begin{aligned}
\mathcal{L}_{\text {recon }} & =\left\|\boldsymbol{I}-\boldsymbol{I}^{\prime}\right\|_{2}^{2}
\end{aligned}
\end{equation}

\begin{figure*}
\centering
\includegraphics[width=1\textwidth]{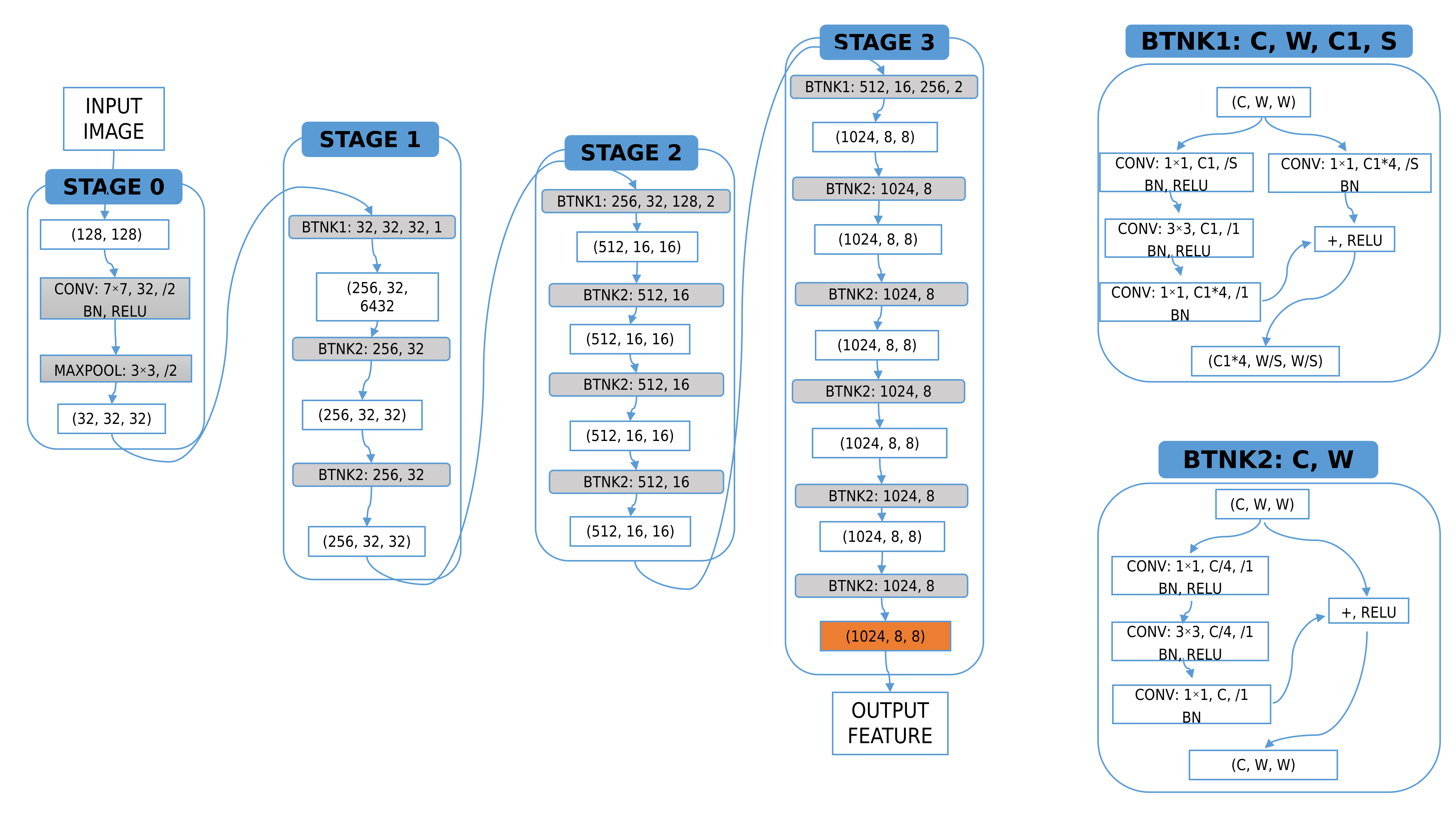}
\caption{Structure of the ResNet as a backbone.}
\label{figure4}
\end{figure*}

\noindent\textbf{Backbone} We employed a pre-trained three-layer ResNet-50 \cite{he2016deep} as the backbone to extract features from the images, as depicted in the Figure \ref{figure4}. In this study, we input two DRRs denoted as $I_1$ and $I_2$, with dimensions of $128\times 128$ pixels each. The backbone is applied individually to extract features from both images, generating two feature blocks of size $8\times8\times1024$. These feature blocks are subsequently linearly mapped to the dimensions of $8\times8\times256$. After concatenating the two feature blocks and incorporating position encoding, represented as $\mathcal{B}_{p}$, we obtain a feature block $t$ with dimensions of $8\times16\times256$, which is then input into the transformer. In this context, $\mathcal{B}$ refers to the backbone.

\begin{equation}
\begin{aligned}
t = [\mathcal{B}({I}),\mathcal{B}(I^{\prime})] +\mathcal{B}_p
\end{aligned}
\end{equation}

\begin{figure*}
\centering
\includegraphics[width=1\textwidth]{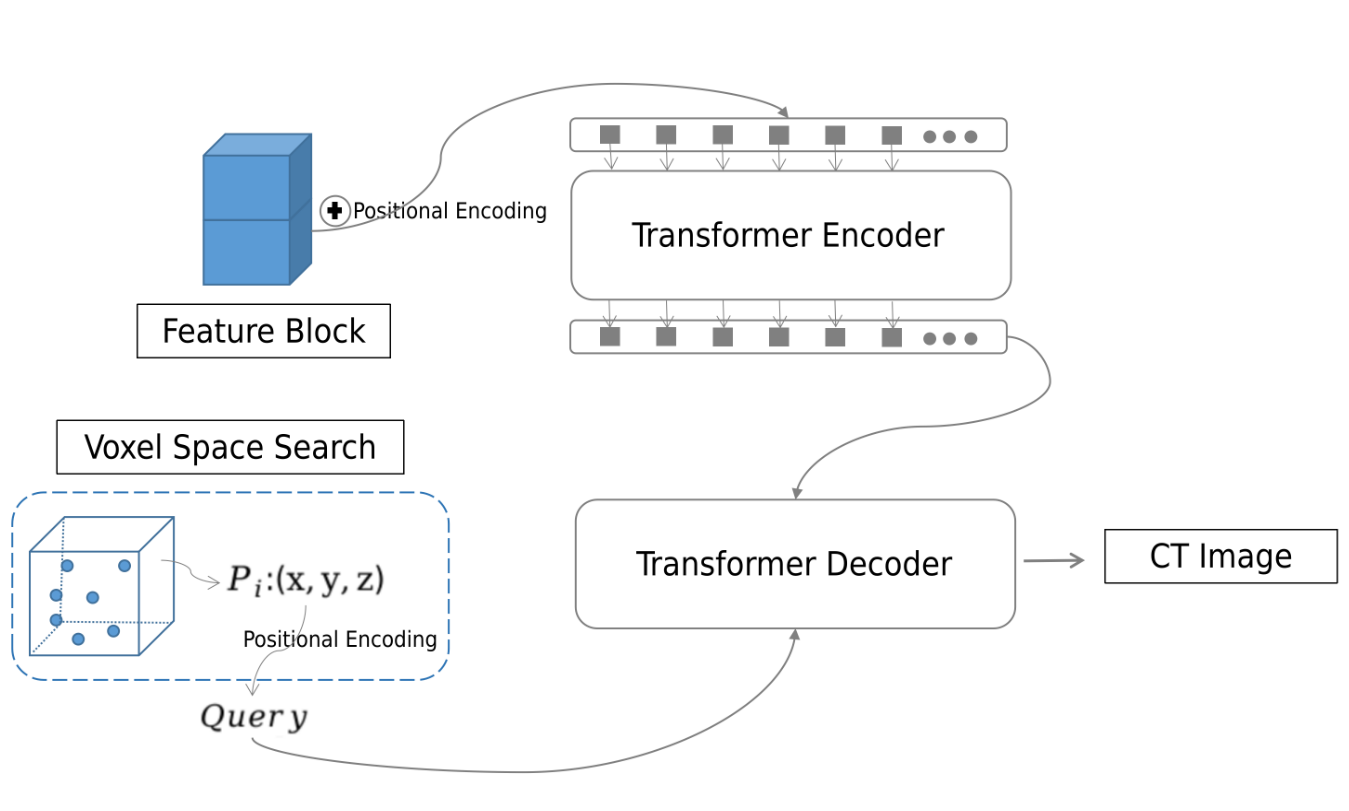} 
\caption{Structure of the transformer.} 
\label{figure5} 
\end{figure*}

%\begin{figure*}
%\centering
%\includegraphics[width=0.7\textwidth]{figure/7.png}
%\caption{Self-attention and cross-attention of two DRR images are input into the network.}
%\label{fig:attention}
%\end{figure*}

\noindent\textbf{Transformer} Our architecture is built upon the Transformer model and utilizes six layers for both the encoder and decoder shown in Figure \ref{figure5}. Each encoder layer is composed of an 8-headed self-attention module, while each decoder layer contains an 8-headed encoder-decoder attention module. As illustrated in the diagram, we input the features $t$ derived from $\mathcal{B}$ into the Transformer encoder $\mathcal{T}_{\mathcal{E}}$, and subsequently input them, alongside the position coordinates and features $\mathcal{V}(\mathcal{P})$, into the Transformer decoder using linear positional encoding mapping. Within the Transformer, features interact and communicate, facilitating information aggregation.

Traditional CNN-based methods, such as X2CT-GAN, face limitations in information aggregation, as they can only capture local information by summing up matrix transpositions acquired through convolution. In contrast, Transformer-based methods provide greater flexibility in information aggregation. In our proposed architecture, we employ self-attention mechanisms to aggregate textual information within an image and cross-modal attention to aggregate information between images and text, resulting in more efficient and effective information aggregation in our model.

\noindent\textbf{MLP} Upon processing the result, the transformer decoder generates a 256-dimensional vector representing the grayscale values of the corresponding coordinates. To regress the grayscale values of these coordinates from the potential vector, we utilize a 3-layer Multilayer Perceptron (MLP). Each layer of the MLP comprises 256 neurons and is activated by the Rectified Linear Unit (ReLU) activation function.

\begin{equation}
\begin{aligned}
\theta^{\prime} = MLP(\mathcal{T}_{D}([\mathcal{T}_\mathcal{E}(t),\mathcal{V}(\mathcal{P})]))
\end{aligned}
\end{equation}

\subsection{Accelerated Voxels Space Search Strategy}
In the previous section, we described the input query for the transformer decoder as $\mathcal{P}:(x,y,z)$, where $\mathcal{P} \in {[0,1]}^{3}$, and the output as a grayscale image. To generate a 128$\times$128$\times$128 3DCTimage,we require 128$\times$128$\times$128 queries uniformly spaced in ${[0,1]}^{3}$. This can be achieved using Meshgrid(0:128:1,0:128:1,0:128:1). However, this approach presents significant challenges in terms of memory consumption and inference time. To address this issue, we adopt an efficient voxel space search strategy in this paper. Specifically, the transformer decoder produces grayscale values $\Theta$ for cubic blocks of size 4$\times$4$\times$4, centered at the respective query coordinates $\mathcal{P}$. This reduces the number of queries required to only 32$\times$32$\times$32, thereby significantly reducing memory consumption and inference time. The network output is given as:

\begin{equation}
\begin{aligned}
\Theta^{\prime} = MLP(\mathcal{T}_{D}([\mathcal{T}_\mathcal{E}(t),\mathcal{V}(\mathcal{P})]))
\end{aligned}
\end{equation}
Thus, the output results $\Theta^{\prime}$ of the network are concatenated to form the final image $I^{\prime}$.

\subsection{Implementation details}
We implemented a two-stage training approach using different training sets in this study. For training our networks, the Adam solver was employed. Initially, we fixed the learning rate of the backbone and trained the transformer based on the pre-trained model with a learning rate of 1e-5. Upon achieving convergence during training, we lowered the learning rate of the backbone to 1e-6 while maintaining the learning rate of the remaining parts at 1e-5. Both training and inference were performed using a batch size of 1 and were conducted on an RTX A6000 48G.

\section{Results $\&$ Discussion}
Considering real-time IGRT as our objective, the evaluation should be based on the task's specific characteristics. First, as it is a generation task, the image quality must be assessed. Second, given the real-time nature of the task, the algorithm's speed must be evaluated. Since the goal is to locate and track the patient's internal conditions, the accuracy of the lung structure must be evaluated through manual delineation. Finally, as the data distribution may differ among hospitals and instruments, the algorithm's generalizability is critical. Thus, we conducted experiments on three datasets to validate the image quality, algorithm speed, structural accuracy, and generalizability of the algorithm.

\subsection{Datasets}

\begin{table*}[]
\label{table:1}
\caption{Comparative experiment on the 50-patients dataset and the LIDC-IDRI Dataset.}
\resizebox{\textwidth}{!}{%
\begin{tabular}{@{}llllllll@{}}
\toprule
\multicolumn{1}{c}{\multirow{2}{*}{Method}} & \multicolumn{3}{c}{50-patients Dataset}                                               & \multicolumn{3}{c}{LIDC-IDRI Dataset}                                                 & \multicolumn{1}{c}{\multirow{2}{*}{time(per image)}} \\ \cmidrule(lr){2-7}
\multicolumn{1}{c}{}                        & \multicolumn{1}{l|}{SSIM} & \multicolumn{1}{l|}{PSNR(dB)} & \multicolumn{1}{l|}{Dice} & \multicolumn{1}{l|}{SSIM} & \multicolumn{1}{l|}{PSNR(dB)} & \multicolumn{1}{l|}{Dice} & \multicolumn{1}{c}{}                                 \\ \midrule
XTransCT                                    & 0.77(0.01)                & 22.1(0.02)                    & 0.91(0.01)                & 0.52(0.01)                & 19.72(0.04)                   & 0.95(0.005)               & 44ms                                                 \\ \midrule
X2CT                                        & 0.70(0.01)                & 21.3(0.03)                    & 0.87(0.01)                & 0.49(0.01)                & 18.16(0.05)                   & 0.94(0.005)               & 163ms                                                \\ \midrule
X2CTGAN                                     & 0.65(0.03)                & 20.5(0.08)                    & 0.83(0.02)                & 0.43(0.02)                & 17.56(0.1)                    & 0.92(0.008)               & 163ms                                                \\ \midrule
2D CNN                                      & 0.46(0.05)                & 18.5(0.08)                    & -                      & 0.29(0.04)                & 15.34(0.3)                    & -                      & 98ms                                                 \\ \midrule

\end{tabular}
}
\end{table*}

\noindent\textbf{Retrospective study dataset from radiotherapy patients.} In this study, we conducted a retrospective analysis of 50 patients who underwent radiotherapy following breast-conserving surgery. Standard treatment planning procedures were employed using CT images obtained with a Siemens Medical System scanner. The voxel dimensions of the CT images were $0.977 \times 0.977 \times 5$ $mm^3$, and the data size was $512 \times 512 \times 80$.

\noindent\textbf{LIDC-IDRI dataset \cite{armato2011lung}} The LIDC-IDRI dataset includes medical imaging data from lung CT scans with multiple nodule images and is sponsored by the National Cancer Institute (NCI) of the United States. This dataset comprises 1,018 medical images, of which 917 were designated for training purposes, and the remaining 101 images were reserved for validation to assess our approach's effectiveness.

\noindent\textbf{LNDb  dataset \cite{pedrosa2019lndb,pedrosa2021lndb}} The LNDb dataset, consisting of 294 CT scans, was retrospectively collected at the Centro Hospitalar e Universitário de São João (CHUSJ) in Porto, Portugal between 2016 and 2018. To comply with ethical standards, the dataset underwent meticulous anonymization, removing all personally identifiable information except for gender and year of birth. The CHUSJ ethical committee approved this process. To verify our algorithm's generalizability, we randomly selected 24 sets of CT scans from the LNDb dataset. We did not train on this specific dataset but instead tested our model trained on the LIDC-IDRI dataset.

\subsection{Metrics}
\begin{figure*}
\centering

\includegraphics[width=1\textwidth]{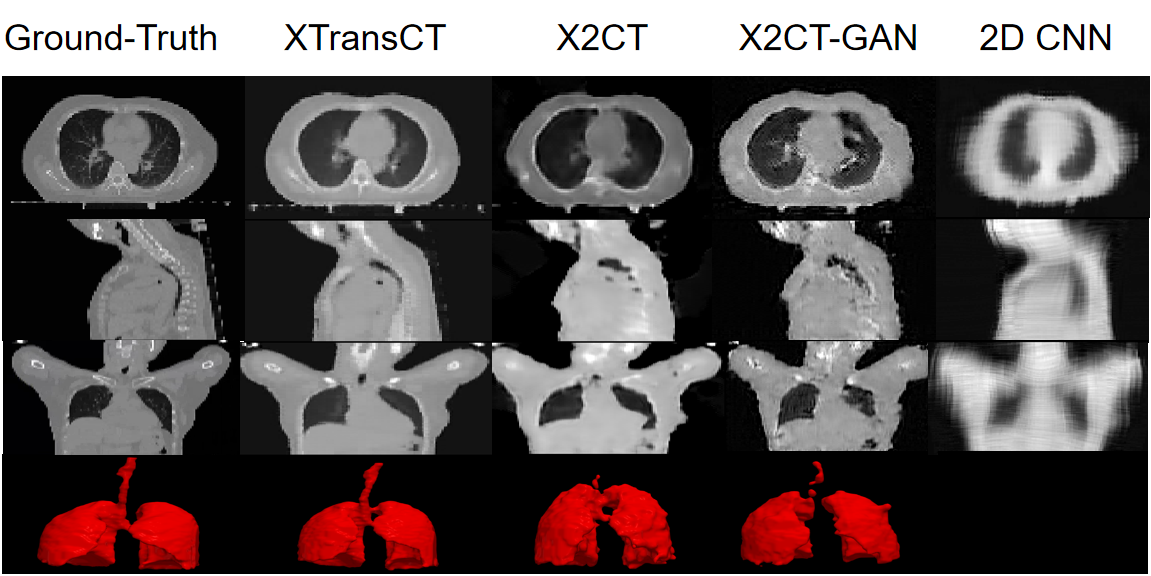}
\caption{Comparative experiment on the 50-patient dataset.
}
\label{fig:Comparative experiment on the 50-patient dataset}
\end{figure*}

\noindent\textbf{Structural Similarity Index (SSIM).}
The Structural Similarity Index (SSIM) is a widely used metric for evaluating digital image quality. It assesses human perception of image quality and measures the similarity and luminance difference between the original and reconstructed images. The SSIM metric provides a numerical value between 0 and 1, where a higher value signifies a greater degree of similarity between the reconstructed image and the original. The formula for SSIM is::

\begin{equation}
\operatorname{SSIM}(x, y)=L(x, y) * c(x, y) * s(x, y)
\end{equation}

Where $L(x, y)$, $c(x, y)$, and $s(x, y)$ represent luminance, contrast, and structural similarity, respectively.

\noindent\textbf{Peak Signal-to-Noise Ratio (PSNR).}
The Peak Signal-to-Noise Ratio (PSNR) is an essential metric for evaluating signal quality across numerous fields, such as computer vision, image processing, and compression algorithms for images and videos. PSNR quantifies the discrepancy between the original and reconstructed signals, expressed in decibels (dB). A higher PSNR value signifies a smaller error between the original and reconstructed signals, thus indicating a higher fidelity of the reconstructed signal to the original. The formula for PSNR is:

\begin{equation}
P S N R=10 * \log 10\left(M A X^2 / M S E\right)
\end{equation}

Where $MAX$ denotes the maximum pixel value and $MSE$ represents the mean squared error. More specifically:

\begin{equation}
M S E=(1 / n) * \sum_{i=1}^n \sum_{j=1}^m\left(I_{i j}-K_{i j}\right)^2
\end{equation}

Here, $n$ and $m$ correspond to the width and height of the images, $I_{ij}$ symbolizes the pixel value in the original image, and $K_{ij}$ signifies the pixel value in the reconstructed image.

\noindent\textbf{Dice coefficient.}
The Dice coefficient is a prevalent measure for assessing segmentation accuracy. In this research, we manually delineated segmentation labels for the generated images and compared them with the ground truth to evaluate the structural accuracy of the generated images. The formula for the Dice coefficient is:

\begin{equation}
\text { Dice }=(2 * T P) /(2 * T P+F P+F N)
\end{equation}

Where $TP$ refers to the overlapping part (true positives), while $FP$ (false positives) and $FN$ (false negatives) represent non-overlapping parts.

\subsection{Experiments}

\begin{figure*}
\includegraphics[width=1\textwidth]{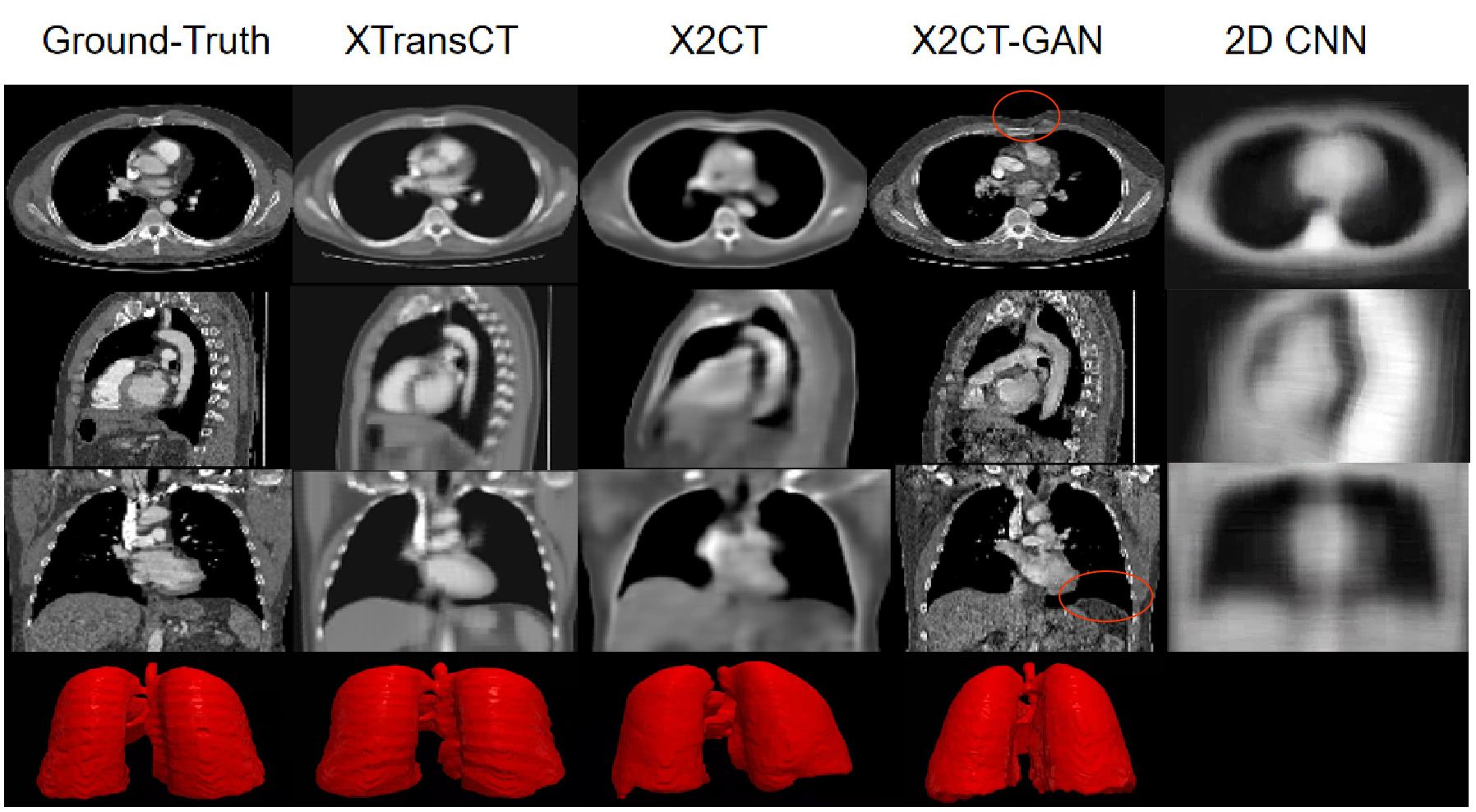}
\centering

\caption{Comparative experiments on the LIDC-IDRI Dataset.
}
\label{fig:LIDC}
\end{figure*}

\begin{figure}
\includegraphics[width=1\textwidth]{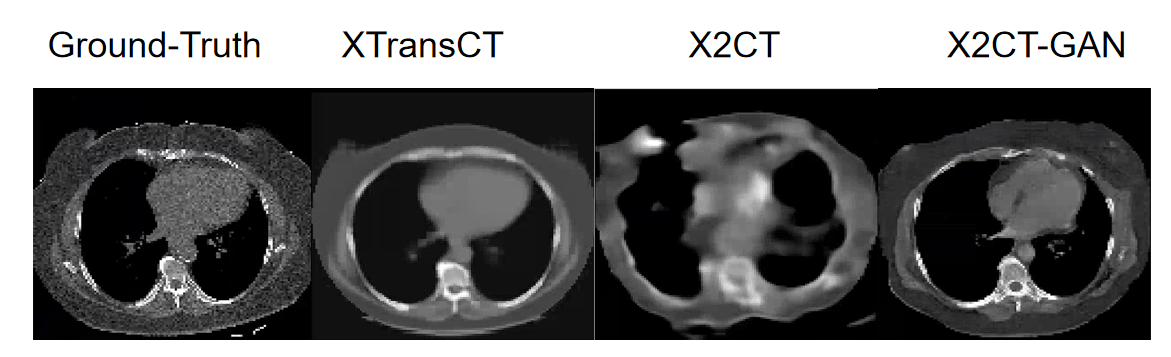}
\centering

\caption{Generalization of the method is verified using comparative experiments on the LNDb Dataset.
}
\label{fig:LNDB}
\end{figure}

Our proposed method's effectiveness and reliability are assessed using two datasets, specifically a small-scale dataset comprising 50 patients and a large-scale publicly available dataset, LIDC-IDRI. We validate our approach on datasets of varying sizes to demonstrate its performance and robustness. For establishing a baseline, we employ X2CT, X2CTGAN, and a 2D CNN, where a 128 $\times$ 128 $\times$ 2 map is converted to a 128 $\times$ 128 $\times$ 128 2D Unet.

Figure \ref{fig:Comparative experiment on the 50-patient dataset} displays the results of our approach on a small-scale dataset of 50 patients. The outcomes show that our method can effectively recover the contour and various features of CT scans. Particularly, the reconstructed images generated by X2CT exhibit relatively blurred details, while X2CTGAN displays severe distortion. We attribute these outcomes to the insufficient data for GANs to learn the distribution accurately. On the other hand, the 2D CNN method fails to reconstruct meaningful boundaries and intricate details. Our manual tracing labels confirm that our method can successfully recover the trachea and lung parenchyma, which are close to the gold standard. In contrast, the trachea and lungs reconstructed by X2CT and X2CT-GAN exhibit fragmentation and distortion. Since the 2D CNN method cannot retrieve the semantic information of 3D CT images, we did not trace the labels for this approach.

The validation results from the LIDC-IDRI dataset are presented in Figure \ref{fig:LIDC}. Our proposed method successfully recovers intricate features of the lungs and bones, yielding images that closely match the ground truth. In contrast, the X2CT method results in significant loss of detail. Although X2CT-GAN produces more realistic images and restores additional details, the reconstructed images appear excessively smoothed compared to our method. However, GAN-based approaches suffer from a drawback of restoring details with errors, as indicated by the red circles in the figure, which is a critical concern for applications such as IGRT.

Quantitative results on two datasets are presented in Table \ref{table:1}. Our approach demonstrates superior performance in terms of image quality (SSIM, PSNR) and structural restoration (Dice) compared to other methods. Moreover, our algorithm exhibits remarkable speed, with a runtime of 44ms, approximately one-fourth that of X2CT-GAN, making our method suitable for real-time IGRT applications.

We conducted a validation of our external dataset validation method using data obtained from the Lung Nodule Database (LNDb). The distributions of datasets from different medical centers and devices often differ, potentially leading to decreased algorithm performance. In this regard, we present the test results of XTransCT, X2CT, and X2CT-GAN, which were trained on the LIDC-LDRI dataset, using the LNDb dataset. As depicted in Figure \ref{fig:LNDB} and summarized in Table \ref{table:2}, our proposed XTransCT and X2CT-GAN models continue to exhibit robust performance, while X2CT shows poor results. We attribute this to the limited ability of the 2D-3D convolution to learn the mapping from two-dimensional to three-dimensional space. In contrast, the GAN constraints facilitate improved outcomes. Furthermore, the embedded attention mechanism in our method strengthens its capacity to learn the 2D-3D mapping, as supported by both quantitative and qualitative outcomes. Consequently, our approach exhibits a robust generalization capability.

\begin{table}[h]
\caption{Comparative Experiment on LNDb Dataset.}
\label{table:2}
\begin{tabularx}{\textwidth}{|l|X|X|X|}
\hline
\multirow{2}{*}{Method} & \multicolumn{3}{c|}{LNDb Dataset} \\ \cline{2-4}
                        & SSIM       & PSNR(dB)   & DICE       \\ \hline
XTransCT                & 0.44(0.01) & 21.0(0.06) & 0.91(0.01) \\ \hline
X2CT                    & 0.26(0.03) & 11.1(0.19) & -          \\ \hline
X2CTGAN                 & 0.41(0.01) & 19.6(0.11) & 0.90(0.01) \\ \hline
\end{tabularx}
\end{table}

\subsection{Discussion}
\noindent\textbf{Use GAN or not?} Although GANs can produce CT images with intricate details and realistic styles, the reliability of the added details remains unverified. In addition, the resulting images' quality, assessed by SSIM and PSNR, and structure, evaluated by DICE, are typically inferior to those generated without GAN. Furthermore, our experimental findings indicate that utilizing GANs may diminish the algorithm's stability, leading to increased variance. Considering our goal of providing guidance for IGRT, this instability is undesirable, and we have opted against employing GAN-based frameworks.

\noindent\textbf{2D CNN + Transformer $>$ 3D CNN.} We introduce an innovative framework that integrates 2D CNN and transformers based on voxel space search. By circumventing the use of computationally-intensive 3D convolutions, our framework achieves a substantial improvement in reconstruction speed compared to prior convolutional frameworks. The transformer-based approach facilitates efficient information aggregation from two images, thereby enhancing our framework's ability to learn the mapping from 2D to 3D space. Our proposed method is validated through improved accuracy and generalization capabilities.

\section{Conclusion}
We propose an innovative framework that combines CNNs and transformers for real-time volumetric CT reconstruction during IGRT. Our research evaluates the image quality, structural accuracy, real-time performance, and generalization capability of the proposed reconstruction method using multi-center datasets of varying sizes, attaining state-of-the-art results.

\section*{Acknowledgments}
This work is partly supported by grants from the National Natural Science Foundation of China (82202954, U20A201795, U21A20480, 12126608) and the Chinese Academy of Sciences Special Research Assistant Grant Program.

\section*{Declarations}
The authors declare that they have no known competing financial interests or personal relationships that could have appeared to influence the work reported in this paper.

\section*{Ethical Statement}
The project was approved by the Ethics Committee of Sun Yat-sen University Cancer Center,with No: SL-G2023-052-01. No animal experiments were involved in this study.

For our medical imaging research, we strictly adhere to the principles embodied in the Helsinki Declaration and local legal requirements. All imaging data were obtained from the hospital's medical imaging database, and patient data have been de-identified to protect their privacy and identity information.

Prior to obtaining the imaging data, all patients (or their legal guardians) have provided written consent for their imaging data to be used for research purposes. In addition, all patients (or their legal guardians) have provided written consent for the publication of research findings, with their identity information not disclosed in the published results.

As this study did not involve clinical trials, a clinical trial registration number cannot be provided.

I hereby confirm the accuracy of the above information and commit to complying with all relevant ethical standards and guidelines.

\bibliography{sn-bibliography}

\begin{thebibliography}{27}
\providecommand{\natexlab}[1]{#1}
\providecommand{\url}[1]{\texttt{#1}}
\expandafter\ifx\csname urlstyle\endcsname\relax
  \providecommand{\doi}[1]{doi: #1}\else
  \providecommand{\doi}{doi: \begingroup \urlstyle{rm}\Url}\fi

\bibitem[Xing et~al.(2006)Xing, Thorndyke, Schreibmann, Yang, Li, Kim, Luxton,
  and Koong]{xing2006overview}
Lei Xing, Brian Thorndyke, Eduard Schreibmann, Yong Yang, Tian-Fang Li, Gwe-Ya
  Kim, Gary Luxton, and Albert Koong.
\newblock Overview of image-guided radiation therapy.
\newblock \emph{Medical Dosimetry}, 31\penalty0 (2):\penalty0 91--112, 2006.

\bibitem[Zhang et~al.(2018)Zhang, Liang, Dong, Xie, and Cao]{zhang2018sparse}
Zhicheng Zhang, Xiaokun Liang, Xu~Dong, Yaoqin Xie, and Guohua Cao.
\newblock A sparse-view ct reconstruction method based on combination of
  densenet and deconvolution.
\newblock \emph{IEEE transactions on medical imaging}, 37\penalty0
  (6):\penalty0 1407--1417, 2018.

\bibitem[Jin et~al.(2008)Jin, Yin, Tenn, Medin, and Solberg]{jin2008use}
Jian-Yue Jin, Fang-Fang Yin, Stephen~E Tenn, Paul~M Medin, and Timothy~D
  Solberg.
\newblock Use of the brainlab exactrac x-ray 6d system in image-guided
  radiotherapy.
\newblock \emph{Medical Dosimetry}, 33\penalty0 (2):\penalty0 124--134, 2008.

\bibitem[Dong et~al.(2023)Dong, Dai, Li, Zhang, He, Liu, Chan, Li, Xie, and
  Liang]{dong20232d}
Guoya Dong, Jingjing Dai, Na~Li, Chulong Zhang, Wenfeng He, Lin Liu, Yinping
  Chan, Yunhui Li, Yaoqin Xie, and Xiaokun Liang.
\newblock 2d/3d non-rigid image registration via two orthogonal x-ray
  projection images for lung tumor tracking.
\newblock \emph{Bioengineering}, 10\penalty0 (2):\penalty0 144, 2023.

\bibitem[Zhang et~al.(2021)Zhang, Xia, Jin, and Gao]{zhang20212d}
Jiayi Zhang, Wei Xia, Qingpeng Jin, and Xin Gao.
\newblock A 2d/3d non-rigid registration method for lung images based on a
  non-linear correlation between displacement vectors and similarity measures.
\newblock \emph{Journal of Medical and Biological Engineering}, 41:\penalty0
  331--342, 2021.

\bibitem[Ying et~al.(2019)Ying, Guo, Ma, Wu, Weng, and Zheng]{ying2019x2ct}
Xingde Ying, Heng Guo, Kai Ma, Jian Wu, Zhengxin Weng, and Yefeng Zheng.
\newblock X2ct-gan: reconstructing ct from biplanar x-rays with generative
  adversarial networks.
\newblock In \emph{Proceedings of the IEEE/CVF conference on computer vision
  and pattern recognition}, pages 10619--10628, 2019.

\bibitem[Sayed et~al.(2022)Sayed, Gibson, Watson, Prisacariu, Firman, and
  Godard]{sayed2022simplerecon}
Mohamed Sayed, John Gibson, Jamie Watson, Victor Prisacariu, Michael Firman,
  and Cl{\'e}ment Godard.
\newblock Simplerecon: 3d reconstruction without 3d convolutions.
\newblock In \emph{Computer Vision--ECCV 2022: 17th European Conference, Tel
  Aviv, Israel, October 23--27, 2022, Proceedings, Part XXXIII}, pages 1--19.
  Springer, 2022.

\bibitem[Mildenhall et~al.(2021)Mildenhall, Srinivasan, Tancik, Barron,
  Ramamoorthi, and Ng]{mildenhall2021nerf}
Ben Mildenhall, Pratul~P Srinivasan, Matthew Tancik, Jonathan~T Barron, Ravi
  Ramamoorthi, and Ren Ng.
\newblock Nerf: Representing scenes as neural radiance fields for view
  synthesis.
\newblock \emph{Communications of the ACM}, 65\penalty0 (1):\penalty0 99--106,
  2021.

\bibitem[Wang et~al.(2017)Wang, Xie, Zhang, Chen, and Gu]{wang2017pulmonary}
Tao Wang, Hongzhi Xie, Shuyang Zhang, Dong Chen, and Lixu Gu.
\newblock A pulmonary deformation registration framework for biplane x-ray and
  ct using sparse motion composition.
\newblock In \emph{2017 IEEE Life Sciences Conference (LSC)}, pages 47--50.
  IEEE, 2017.

\bibitem[Ge et~al.(2022)Ge, He, Xia, Xu, Sun, Yang, Li, Wang, Yu, Zhang,
  et~al.]{ge2022x}
Rongjun Ge, Yuting He, Cong Xia, Chenchu Xu, Weiya Sun, Guanyu Yang, Junru Li,
  Zhihua Wang, Hailing Yu, Daoqiang Zhang, et~al.
\newblock X-ctrsnet: 3d cervical vertebra ct reconstruction and segmentation
  directly from 2d x-ray images.
\newblock \emph{Knowledge-Based Systems}, 236:\penalty0 107680, 2022.

\bibitem[Zhang et~al.(2016)Zhang, Tehrani, and Wang]{ref13}
You Zhang, Joubin~Nasehi Tehrani, and Jing Wang.
\newblock A biomechanical modeling guided cbct estimation technique.
\newblock \emph{IEEE transactions on medical imaging}, 36\penalty0
  (2):\penalty0 641--652, 2016.

\bibitem[Wei et~al.(2019)Wei, Zhou, Liu, Bai, Fu, Li, Liang, and Wu]{ref19}
Ran Wei, Fugen Zhou, Bo~Liu, Xiangzhi Bai, Dongshan Fu, Yongbao Li, Bin Liang,
  and Qiuwen Wu.
\newblock Convolutional neural network (cnn) based three dimensional tumor
  localization using single x-ray projection.
\newblock \emph{IEEE Access}, 7:\penalty0 37026--37038, 2019.

\bibitem[Ketcha et~al.(2017)Ketcha, De~Silva, Uneri, Jacobson, Goerres,
  Kleinszig, Vogt, Wolinsky, and Siewerdsen]{ref18}
MD~Ketcha, T~De~Silva, A~Uneri, MW~Jacobson, J~Goerres, G~Kleinszig, S~Vogt,
  JP~Wolinsky, and JH~Siewerdsen.
\newblock Multi-stage 3d--2d registration for correction of anatomical
  deformation in image-guided spine surgery.
\newblock \emph{Physics in Medicine \& Biology}, 62\penalty0 (11):\penalty0
  4604, 2017.

\bibitem[Gao et~al.(2020)Gao, Liu, Gu, Killeen, Armand, Taylor, and
  Unberath]{ref15}
Cong Gao, Xingtong Liu, Wenhao Gu, Benjamin Killeen, Mehran Armand, Russell
  Taylor, and Mathias Unberath.
\newblock Generalizing spatial transformers to projective geometry with
  applications to 2d/3d registration.
\newblock In \emph{International Conference on Medical Image Computing and
  Computer-Assisted Intervention}, pages 329--339. Springer, 2020.

\bibitem[Liao et~al.(2019)Liao, Lin, Zhang, Zhang, Luo, and Zhou]{ref16}
Haofu Liao, Wei-An Lin, Jiarui Zhang, Jingdan Zhang, Jiebo Luo, and S.~Kevin
  Zhou.
\newblock Multiview 2d/3d rigid registration via a point-of-interest network
  for tracking and triangulation.
\newblock In \emph{2019 IEEE/CVF Conference on Computer Vision and Pattern
  Recognition (CVPR)}, pages 12630--12639, 2019.
\newblock \doi{10.1109/CVPR.2019.01292}.

\bibitem[Li et~al.(2020)Li, Pei, Guo, Ma, Xu, and Zha]{ref25}
Peixin Li, Yuru Pei, Yuke Guo, Gengyu Ma, Tianmin Xu, and Hongbin Zha.
\newblock Non-rigid 2d-3d registration using convolutional autoencoders.
\newblock In \emph{2020 IEEE 17th International Symposium on Biomedical Imaging
  (ISBI)}, pages 700--704. IEEE, 2020.

\bibitem[Park et~al.(2019)Park, Florence, Straub, Newcombe, and
  Lovegrove]{park2019deepsdf}
Jeong~Joon Park, Peter Florence, Julian Straub, Richard Newcombe, and Steven
  Lovegrove.
\newblock Deepsdf: Learning continuous signed distance functions for shape
  representation.
\newblock In \emph{Proceedings of the IEEE/CVF conference on computer vision
  and pattern recognition}, pages 165--174, 2019.

\bibitem[Jiang et~al.(2021)Jiang, Trulls, Hosang, Tagliasacchi, and
  Yi]{jiang2021cotr}
Wei Jiang, Eduard Trulls, Jan Hosang, Andrea Tagliasacchi, and Kwang~Moo Yi.
\newblock Cotr: Correspondence transformer for matching across images.
\newblock In \emph{Proceedings of the IEEE/CVF International Conference on
  Computer Vision}, pages 6207--6217, 2021.

\bibitem[Jaderberg et~al.(2015)Jaderberg, Simonyan, Zisserman,
  et~al.]{jaderberg2015spatial}
Max Jaderberg, Karen Simonyan, Andrew Zisserman, et~al.
\newblock Spatial transformer networks.
\newblock \emph{Advances in neural information processing systems}, 28, 2015.

\bibitem[Vaswani et~al.(2017)Vaswani, Shazeer, Parmar, Uszkoreit, Jones, Gomez,
  Kaiser, and Polosukhin]{vaswani2017attention}
Ashish Vaswani, Noam Shazeer, Niki Parmar, Jakob Uszkoreit, Llion Jones,
  Aidan~N Gomez, {\L}ukasz Kaiser, and Illia Polosukhin.
\newblock Attention is all you need.
\newblock \emph{Advances in neural information processing systems}, 30, 2017.

\bibitem[Carion et~al.(2020)Carion, Massa, Synnaeve, Usunier, Kirillov, and
  Zagoruyko]{carion2020end}
Nicolas Carion, Francisco Massa, Gabriel Synnaeve, Nicolas Usunier, Alexander
  Kirillov, and Sergey Zagoruyko.
\newblock End-to-end object detection with transformers.
\newblock In \emph{Computer Vision--ECCV 2020: 16th European Conference,
  Glasgow, UK, August 23--28, 2020, Proceedings, Part I 16}, pages 213--229.
  Springer, 2020.

\bibitem[Dosovitskiy et~al.(2020)Dosovitskiy, Beyer, Kolesnikov, Weissenborn,
  Zhai, Unterthiner, Dehghani, Minderer, Heigold, Gelly,
  et~al.]{dosovitskiy2020image}
Alexey Dosovitskiy, Lucas Beyer, Alexander Kolesnikov, Dirk Weissenborn,
  Xiaohua Zhai, Thomas Unterthiner, Mostafa Dehghani, Matthias Minderer, Georg
  Heigold, Sylvain Gelly, et~al.
\newblock An image is worth 16x16 words: Transformers for image recognition at
  scale.
\newblock \emph{arXiv preprint arXiv:2010.11929}, 2020.

\bibitem[Liu et~al.(2021)Liu, Lin, Cao, Hu, Wei, Zhang, Lin, and
  Guo]{liu2021swin}
Ze~Liu, Yutong Lin, Yue Cao, Han Hu, Yixuan Wei, Zheng Zhang, Stephen Lin, and
  Baining Guo.
\newblock Swin transformer: Hierarchical vision transformer using shifted
  windows.
\newblock In \emph{Proceedings of the IEEE/CVF international conference on
  computer vision}, pages 10012--10022, 2021.

\bibitem[He et~al.(2016)He, Zhang, Ren, and Sun]{he2016deep}
Kaiming He, Xiangyu Zhang, Shaoqing Ren, and Jian Sun.
\newblock Deep residual learning for image recognition.
\newblock In \emph{Proceedings of the IEEE conference on computer vision and
  pattern recognition}, pages 770--778, 2016.

\bibitem[Armato~III et~al.(2011)Armato~III, McLennan, Bidaut, McNitt-Gray,
  Meyer, Reeves, Zhao, Aberle, Henschke, Hoffman, et~al.]{armato2011lung}
Samuel~G Armato~III, Geoffrey McLennan, Luc Bidaut, Michael~F McNitt-Gray,
  Charles~R Meyer, Anthony~P Reeves, Binsheng Zhao, Denise~R Aberle, Claudia~I
  Henschke, Eric~A Hoffman, et~al.
\newblock The lung image database consortium (lidc) and image database resource
  initiative (idri): a completed reference database of lung nodules on ct
  scans.
\newblock \emph{Medical physics}, 38\penalty0 (2):\penalty0 915--931, 2011.

\bibitem[Pedrosa et~al.(2019)Pedrosa, Aresta, Ferreira, Rodrigues, Leit{\~a}o,
  Carvalho, Rebelo, Negr{\~a}o, Ramos, Cunha, et~al.]{pedrosa2019lndb}
Jo{\~a}o Pedrosa, Guilherme Aresta, Carlos Ferreira, M{\'a}rcio Rodrigues,
  Patr{\'\i}cia Leit{\~a}o, Andr{\'e}~Silva Carvalho, Jo{\~a}o Rebelo, Eduardo
  Negr{\~a}o, Isabel Ramos, Ant{\'o}nio Cunha, et~al.
\newblock Lndb: a lung nodule database on computed tomography.
\newblock \emph{arXiv preprint arXiv:1911.08434}, 2019.

\bibitem[Pedrosa et~al.(2021)Pedrosa, Aresta, Ferreira, Atwal, Phoulady, Chen,
  Chen, Li, Wang, Galdran, et~al.]{pedrosa2021lndb}
Jo{\~a}o Pedrosa, Guilherme Aresta, Carlos Ferreira, Gurraj Atwal, Hady~Ahmady
  Phoulady, Xiaoyu Chen, Rongzhen Chen, Jiaoliang Li, Liansheng Wang, Adrian
  Galdran, et~al.
\newblock Lndb challenge on automatic lung cancer patient management.
\newblock \emph{Medical image analysis}, 70:\penalty0 102027, 2021.

\end{thebibliography}
\end{document}